# The Imbalanced User-AI Relationships as an Ethical Failure of Front-End Design in Healthcare AI

Imbalanced User-AI Relationships in Healthcare AI

MAUREEN MGHAMBI MWADIME

School of Computer Science, Northumbria University, Newcastle - United Kingdom,

maureen.mwadime@northumbria.ac.uk

**ABSTRACT**

Artificial Intelligence (AI) systems mediate healthcare decisions influencing diagnosis, treatment pathways and patient engagement. While there have been discussions around ethical AI in healthcare focusing on back-end concerns such as accountability, bias and explainability, the front-end interface remains under discussed although it is a critical site where ethical failures arise[1,11]. Adding to the discourse on "Ethics at the Front-End" this paper identifies imbalanced user-AI relationship as a distinct class of front-end ethical failure: one in which users such as patients are rendered highly visible to AI systems through data collection and inference, while remaining unable to understand, interpret or respond to how they are represented. The paper positions this imbalance as an ethical failure of front-end design that undermines users' agency, meaningful participation and effective human oversight, even where AI systems are technically accurate[12,17]. Through the concept of asymmetric legibility and a case illustration of chat-based telemedicine, it demonstrates how specific design choices including default recommendations, restricted input mechanisms and suppressed uncertainty, produce and sustain these imbalances. Focusing on the healthcare context, it further discusses how such choices shape accountability, clinician authority and patient trust. In response, it proposes reciprocity as a design orientation and contributes concrete front-end design interventions that demonstrate how more balanced, participatory user-AI relationships can be operationalized in practice.

**ACM Classification Keywords**

Information systems ⟶ Information systems applications ⟶ Decision support systems ⟶ Expert systems

Additional Keywords and Phrases: User-AI relationships, imbalance, front-end design, ethics, healthcare AI, human oversight

## 1 INTRODUCTION

AI systems are embedded in healthcare, shaping a broad spectrum of activities such as clinical decision making, risk stratification, triage and patient-facing services[4]. These systems promise efficiency, consistency and improved outcomes, thus holding the potential to transform healthcare delivery[5]. However, these technological advancements also fundamentally alter how patients experience their right to health and engage with care processes. In practice, patients rarely encounter AI systems directly; instead, they meet AI through front-end interfaces that present recommendations, risk scores or default care pathways[1,2]. To the patients, it is at this interface – not in the algorithm – that their right to health is realized.

The bulk of ethical AI discourse has predominantly focused on back-end challenges such as mitigating bias, ensuring fairness, promoting transparency and enabling auditability[1,2,11]. While these concerns are essential for trustworthy AI, they risk overlooking the ethical tensions that emerge most acutely in the front-end interface where patients and clinicians interact with AI outputs. Provision of ethical healthcare is not solely defined by the technical accuracy of decisions, but also how those decisions are communicated, negotiated and experienced by the people involved[15]. Established clinical norms including informed consent, shared decision-making and respect for patient voice should not be optional in AI-enhanced contexts[3,15].

The above concern is partly structural. Not all users of healthcare AI systems will participate in back-end design or algorithm development, nor should they be expected to[11]. The front end, however, is where every class of user e.g., patient, clinician, administrator, exercises agency at the moment of interaction[1,13]. Front-end design choices therefore carry significant ethical weight: they determine whether patients and clinicians can engage in meaningful dialogue, exercise independent judgment and hold AI-mediated decisions to account. Attending to these choices is not a secondary concern; it is central to the ethical integrity of AI-enhanced healthcare. This paper argues that front-end interface design must be treated as a site of ethical practice in its own right - one that requires principled frameworks capable of preserving patient agency, supporting clinician judgment and ensuring that AI-mediated care remains accountable to the people it serves.

## 2  HOW FRONT-END DESIGN PRODUCES IMBALANCED USER-AI RELATIONSHIPS

Front-end design is not a neutral conduit for AI outputs - it actively shapes who holds interpretive authority, who can act and who is expected to comply. In many AI-enhanced healthcare interfaces, users are often unable to ask why a recommendation was made, challenge system outputs, or contribute missing contextual information[2,11]. The result is an interaction structure that is built around compliance rather than participation, in direct tension with longstanding ethical expectations of shared decision-making in healthcare[2,3,9].

One of the most consequential patterns driving this imbalance is the framing of AI-generated recommendations as authoritative. Visual prominence, default selections and prescriptive language, position system outputs as primary decision drivers, while reducing patient and clinician input to a function of confirmation or acceptance[8]. Consider a clinical decision-support tool that displays a medication recommendation in bold at the top of the screen with the prescribing option pre-ticked. The clinician must actively untick it to deviate. While this may be viewed by some as a "small" design choice that frames agreement as the default and disagreement as the exception. Furthermore, HCI research has demonstrated that such patterns elevate system authority over human judgment, increasing susceptibility to automation bias: the tendency to over-rely on automated outputs even when uncertainty or contextual mismatch warrants greater scrutiny [8,9]. In clinical contexts, the blurring of critical engagement weakens professional accountability and reduces the capacity of clinicians to exercise the contextual judgment that AI systems cannot replicate[4,16].

Interaction imbalances are also compounded by restricted input mechanisms. Most front-end systems confine users to predefined options or scripted prompts, barring the user's ability to communicate contextual knowledge, uncertainty or lived experience that falls outside system categories. This constraint occurs at the level of interface architecture - through input fields, prompt structures and conversational flows that predefine what counts as valid information. While such guardrails may serve legitimate goals of efficiency and standardisation, they simultaneously narrow the epistemic space of the interaction, silencing forms of knowledge that may be significant[6].

The ethical implications of these design-driven imbalances extend beyond individual experiences. When patients cannot understand how their data is interpreted or interrogate the basis of a recommendation, their capacity to participate meaningfully in healthcare decisions is structurally compromised[3]. When clinicians are presented with AI outputs that appear authoritative but resist interrogation or contextualization, their capacity to exercise professional judgment is undermined. Taken together, these dynamics risk transferring effective decision-making authority from human actors to AI systems - a shift with serious consequences for accountability, trust and the legitimacy of AI-mediated care [16].

## 3 CASE ILLUSTRATION: CHAT-BASED TELEMEDICINE AND ASYMMETRIC LEGIBILITY

A useful concept for examining front-end design failures is asymmetric legibility which arises when an AI system can read and act on user inputs, but users cannot fully interrogate the system's reasoning in return[7]. This is not a transparency failure in the back-end sense of unexplainable algorithms; it is a front-end design choice about what to surface, to whom and when.

Consider a chat-based telemedicine platform deployed in Africa, in which an AI system suggests responses and proposed care steps directly within the shared clinician-patient interface. The interface offers no indication of how suggestions are generated, which patient inputs are prioritised or what information might be missing. The system renders the patient legible - their symptoms and responses are processed and acted upon - while remaining largely illegible to both patient and clinician in return.

This asymmetry constrains participation at two critical points. At input, patients are limited to predefined prompt options and cannot introduce contextual nuance, express uncertainty in their own words or correct misreadings of their situation - they are knowable to the system but unable to engage it on their own terms. At output, recommendations are presented without mechanisms for questioning, adjustment or negotiation; clinicians can technically override suggestions but are given no scaffolding to interrogate or contextualise them within the consultation flow, making meaningful oversight nominal rather than substantive. From a front-end ethics perspective, this constitutes bad design practice: interaction structures that prioritise efficiency and system authority at the expense of participation and interpretability.

This case is not unique to its specific deployment context. Similar chat-based health applications across the globe constrain user agency through limited input options and opaque outputs[2,14]. This undermines oversight in several ways. First, patients are excluded as oversight actors despite being directly affected by AI-mediated decisions. Second, clinicians encounter outputs that appear authoritative and resistant to interrogation, encouraging automation bias and discouraging critical engagement. Third, responsibility for AI-mediated decisions becomes insufficiently attributable to identifiable human decision-makers, weakening accountability and meaningful oversight.

## 4 DESIGNING TOWARDS BALANCED USER-AI RELATIONSHIPS AT THE FRONT-END

To address the ethical failures identified above, this paper proposes reciprocity as a design orientation for front-end AI in healthcare - a principle that positions the interface not as a delivery mechanism for AI outputs, but as a space of exchange in which users can interrogate, contextualise and meaningfully respond to system reasoning. Reciprocity does not require patients to grasp technical or algorithmic details; it demands interfaces that support mutual intelligibility, where systems can render patients legible, and patients and clinicians can in turn see and understand how systems represent and interpret their data[1,14]. Within this orientation, AI outputs are treated as provisional contributions to a clinical dialogue rather than final decisions, and deviation from those outputs is made visible and legitimate rather than discouraged.

In practice, this translates into concrete design choices. Where current interfaces present a single recommendation as a default without explanation, a reciprocal design would surface an expandable rationale - a "why this recommendation?" layer that makes the factors influencing a suggestion visible and contestable[1]. Where fixed prompts restrict users to system-defined categories, free-text fields and explicit "add context" options would enable patients to contribute narrative and experiential knowledge that falls outside those categories[1]. Where the clinician's ability to deviate from an AI suggestion is hidden or absent, a visible and workflow-integrated override function would normalise critical engagement as an expected part of the interaction; and where uncertainty is suppressed in favour of authoritative outputs, confidence indicators and alternative care pathways would give both patients and clinicians the information necessary to exercise judgment rather than defer to it[1].

Human-centred AI research supports this approach, emphasising the importance of interfaces that foreground uncertainty, enable explanation and interrogation, and make AI representations both visible and revisable[1,10,12]. Applied to healthcare, this means interfaces: that allow patients and clinicians to view, question and correct inferred attributes or classifications; that acknowledge system limitations transparently rather than presenting probabilistic outputs as definitive facts; and that preserve the conditions for dialogue and choice consistent with offline standards of care - including mechanisms for expressing disagreement, seeking clarification or refusing AI-mediated recommendations[1,7,12].

These interactional features do not diminish the utility of AI-generated recommendations, nor do they undermine clinical authority or system performance. Rather, they situate those recommendations within an interaction structure that restores the relational qualities fundamental to ethical healthcare - supporting trust, accountability and meaningful human oversight at the point of interaction[3,17].

## 5 CONCLUSION

Ethical failures in AI-mediated healthcare extend beyond biased models or inaccurate predictions; they also emerge at the front end, where patients and clinicians are rendered legible to systems that remain opaque in return, and where responsibility for AI-mediated decisions becomes difficult to attribute to identifiable human actors. Imbalanced user-AI relationships therefore constitute a distinct class of ethical failure, with direct implications for participation, accountability and oversight that cannot be resolved through back-end technical interventions alone.

This has significant implications for how human oversight is understood in governance and policy. A system may satisfy formal transparency requirements through disclosure mechanisms yet still fail to support meaningful oversight in practice, if users cannot understand, question or influence its outputs at the point of interaction. While emerging frameworks such as the EU AI Act explicitly require meaningful human control over high-risk AI systems, this is yet to be fully operationalised at the front-end[17].

From a human-centred perspective, ethical AI in healthcare must therefore begin not just with prediction accuracy, but with interfaces that uphold agency, enable participation and make human oversight genuinely possible at the point of interaction. The design orientation proposed in this paper - reciprocity - offers one such starting point: a principle that re-centres the front end as a site of ethical practice and holds interface design accountable to the same standards of informed consent, shared decision-making and respect for patient voice that have long defined good and quality healthcare service provision.

**References**

[1] Saleema Amershi, Dan Weld, Mihaela Vorvoreanu, Adam Fourney, Besmira Nushi, Penny Collisson, Jina Suh, Shamsi Iqbal, Paul N. Bennett, Kori

Inkpen, Jaime Teevan, Ruth Kikin-Gil, and Eric Horvitz. 2019. Guidelines for Human-AI Interaction. In Proceedings of the 2019 CHI Conference on Human Factors in Computing Systems (CHI '19), 1–13. https://doi.org/10.1145/3290605.3300233

[2] Mike Ananny and Kate Crawford. 2018. Seeing without knowing: Limitations of the transparency ideal and its application to algorithmic accountability. New Media & Society 20, 3: 973–989. https://doi.org/10.1177/1461444816676645

[3] Tom Beauchamp and James Childress. 2019. Principles of Biomedical Ethics: Marking Its Fortieth Anniversary. The American Journal of Bioethics 19, 11: 9–12. https://doi.org/10.1080/15265161.2019.1665402

[4] Robert Challen, Joshua Denny, Martin Pitt, Luke Gompels, Tom Edwards, and Krasimira Tsaneva-Atanasova. 2019. Artificial intelligence, bias and clinical safety. BMJ Quality & Safety 28, 3: 231–237. https://doi.org/10.1136/bmjqs-2018-008370

[5] Polat Goktas, Andrzej Grzybowski, Polat Goktas, and Andrzej Grzybowski. 2025. Shaping the Future of Healthcare: Ethical Clinical Challenges and Pathways to Trustworthy AI. Journal of Clinical Medicine 14, 5. https://doi.org/10.3390/jcm14051605

[6] Donna Haraway. 1988. Situated Knowledges: The Science Question in Feminism and the Privilege of Partial Perspective. Feminist Studies 14, 3: 575. https://doi.org/10.2307/3178066

[7] Joseph Lindley, Haider Ali Akmal, Franziska Pilling, and Paul Coulton. 2020. Researching AI Legibility through Design. In Proceedings of the 2020 CHI Conference on Human Factors in Computing Systems (CHI '20), 1–13. https://doi.org/10.1145/3313831.3376792

[8] Raja Parasuraman and Dietrich H. Manzey. 2010. Complacency and Bias in Human Use of Automation: An Attentional Integration. Human Factors 52, 3: 381–410. https://doi.org/10.1177/0018720810376055

[9] Raja Parasuraman and Victor Riley. 1997. Humans and Automation: Use, Misuse, Disuse, Abuse. Human Factors 39, 2: 230–253. https://doi.org/10.1518/001872097778543886

[10] Stefan Schmager, Ilias O. Pappas, and Polyxeni Vassilakopoulou. 2025. Understanding Human-Centred AI: a review of its defining elements and a research agenda. Behaviour & Information Technology 44, 15: 3771–3810. https://doi.org/10.1080/0144929X.2024.2448719

[11] Andrew D. Selbst, Danah Boyd, Sorelle A. Friedler, Suresh Venkatasubramanian, and Janet Vertesi. 2019. Fairness and Abstraction in Sociotechnical Systems. In Proceedings of the Conference on Fairness, Accountability, and Transparency (FAT* '19), 59–68. https://doi.org/10.1145/3287560.3287598

[12] Ben Shneiderman. 2020. Human-Centered Artificial Intelligence: Reliable, Safe & Trustworthy. https://doi.org/10.48550/arXiv.2002.04087

[13] Lucy Suchman. Human–Machine Reconfigurations: Plans and Situated Actions, 2nd Edition.

[14] Tao Tu, Mike Schaekermann, Anil Palepu, Khaled Saab, Jan Freyberg, Ryutaro Tanno, Amy Wang, Brenna Li, Mohamed Amin, Yong Cheng, Elahe Vedadi, Nenad Tomasev, Shekoofeh Azizi, Karan Singhal, Le Hou, Albert Webson, Kavita Kulkarni, S. Sara Mahdavi, Christopher Semturs, Juraj Gottweis, Joelle Barral, Katherine Chou, Greg S. Corrado, Yossi Matias, Alan Karthikesalingam, and Vivek Natarajan. 2025. Towards conversational diagnostic artificial intelligence. Nature 642, 8067: 442–450. https://doi.org/10.1038/s41586-025-08866-7

[15] Basil Varkey. 2021. Principles of Clinical Ethics and Their Application to Practice. Medical Principles and Practice 30, 1: 17–28. https://doi.org/10.1159/000509119

[16] Ellison B. Weiner, Irene Dankwa-Mullan, William A. Nelson, and Saeed Hassanpour. 2025. Ethical challenges and evolving strategies in the integration of artificial intelligence into clinical practice. PLOS Digital Health 4, 4: e0000810. https://doi.org/10.1371/journal.pdig.0000810

[17] 2024. Regulation (EU) 2024/1689 of the European Parliament and of the Council of 13 June 2024 laying down harmonised rules on artificial intelligence and amending Regulations (EC) No 300/2008, (EU) No 167/2013, (EU) No 168/2013, (EU) 2018/858, (EU) 2018/1139 and (EU) 2019/2144 and Directives 2014/90/EU, (EU) 2016/797 and (EU) 2020/1828 (Artificial Intelligence Act) (Text with EEA relevance). Retrieved December 10, 2025 from http://data.europa.eu/eli/reg/2024/1689/oj